\documentclass[twocolumn,secnumarabic,amssymb, nobibnotes, aps, prl,superscriptaddress,floatfix]{revtex4-2}
\pdfoutput=1

\usepackage{graphicx}
\usepackage{gensymb}
\usepackage{wrapfig}
\usepackage{amsmath}
\usepackage{xcolor}
\usepackage{physics}
\usepackage{siunitx}
\usepackage{upgreek}
\usepackage{braket}
\usepackage[colorlinks,linkcolor=black,citecolor=black,urlcolor=black,filecolor=black]{hyperref}
\begin{document}

\title{Patterning programmable spin arrays on DNA origami for quantum technologies}

\author{Zhiran Zhang}
\thanks{These authors contributed equally}
\affiliation{Department of Physics, University of California, Santa Barbara, California 93106, USA}
\author{Taylor Morrison}
\thanks{These authors contributed equally}
\affiliation{Department of Physics, University of California, Santa Barbara, California 93106, USA}
\author{Lillian Hughes}
\affiliation{Materials Department, University of California, Santa Barbara, California 93106, USA}
\author{Weijie Wu}
\affiliation{Department of Physics, Harvard University, Cambridge, Massachusetts 02138, USA}
\author{Ruiyao Liu}
\affiliation{Department of Physics, University of California, Santa Barbara, California 93106, USA}
\author{Dolev Bluvstein}
\affiliation{Department of Physics, Harvard University, Cambridge, Massachusetts 02138, USA}
\affiliation{Physics Department, California Institute of Technology, Pasadena, California, 91125, USA}
\author{Norman Yao}
\affiliation{Department of Physics, Harvard University, Cambridge, Massachusetts 02138, USA}
\author{Deborah Fygenson}
\affiliation{Department of Physics, University of California, Santa Barbara, California 93106, USA}
\author{Ania C. Bleszynski Jayich}
\email{ania@physics.ucsb.edu}
\affiliation{Department of Physics, University of California, Santa Barbara, California 93106, USA}

\begin{abstract}
The controlled assembly of solid-state spins with nanoscale spatial precision is an outstanding challenge for quantum technology. Here, we combine DNA-based patterning with nitrogen-vacancy (NV) ensemble quantum sensors in diamond to form and sense programmable 2D arrays of spins. We use DNA origami to control the spacing of chelated Gd$^{3+}$ spins, as verified by the observed linear relationship between proximal NVs' relaxation rate, $1/T_1$, and the engineered number of Gd$^{3+}$ spins per origami unit. We further show that DNA origami provides a robust way of functionalizing the diamond surface with spins as it preserves the charge state and spin coherence of proximal, shallow NV centers.   Our work enables the formation and interrogation of ordered, strongly interacting spin networks with applications in quantum sensing and quantum simulation. We quantitatively discuss the prospects of entanglement-enhanced metrology and high-throughput proteomics.

\end{abstract}

\maketitle

\def\thefootnote{*}\footnotetext{These authors contributed equally to this work}\def\thefootnote{\arabic{footnote}}

\setcounter{secnumdepth}{1}
\section{Introduction}

Deterministically positioning individual atomic-scale qubits at the nanoscale provides a means for controlling their interactions, but remains a major challenge for solid-state quantum technologies. At close enough proximity, the dipolar coupling between solid-state or molecular qubits can play a major role in their dynamics; 
tuning the strength and homogeneity of these interactions underpins the development of local quantum memories \cite{Childress_Hanson_2013,Bradley_2019_register,Zaiser2016,Pfender2017}, entanglement-enhanced quantum sensors \cite{Cai2013,Block2024,wu_squeezing_2025,gao_squeezing_2025}, and multi-qubit sensors \cite{Sushkov2014,Tianxing2022,zhiran2023} with improved sensitivity and access to spatio-temporal correlations \cite{Jared2022,cheng_massively_2025}. This motivates the search for a versatile assembly platform that offers precise spatial control, high fidelity, and scalability for engineering interacting spin systems.

Several approaches have been developed to position individual solid-state spin defects at deterministic locations within crystals, ranging from masked ion implantation~\cite{Mengqi2022} to lithographically controlled laser and electron irradiation \cite{Enke_2025_lasercreation,Sunghoon2025,Claire2016}. 
While some of these methods can generate single isolated defects, the spatial resolution of these techniques is typically limited to $\sim$ 10-100 nm, constrained by optical diffraction or by vacancy diffusion during subsequent annealing. Moreover, the generation of extraneous defects during these processes can introduce unwanted decoherence sources. For molecular spin qubits, metal-organic frameworks (MOFs) offer one modular and chemically tunable route for organizing spin arrays. However, while MOFs can organize spin centers into extended periodic lattices, they lack site-specific programmability: one cannot designate which molecular site hosts a particular qubit species, nor create arbitrary, user-defined spatial patterns without redesigning the entire framework’s coordination geometry \cite{vujevic_improving_2023}.  
Alternatively, individual adatom spins can be precisely arranged using scanning tunneling microscopy (STM) \cite{eigler_stm_1990}, which affords exceptional spatial control but suffers from poor scalability and demands both cryogenic conditions and substantial experimental overhead.

\begin{figure}[b]

\includegraphics[width=86mm]
{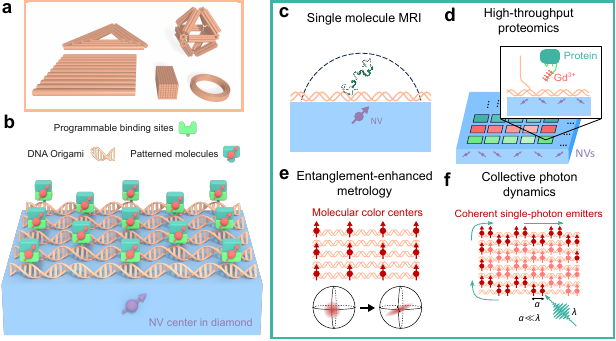}
\centering
\caption{\label{fig:1} \textbf{DNA origami as a programmable platform for patterning spins and interfacing with NV centers.} (a) DNA origami provides a bottom-up structural framework with nanometer-scale spatial control and diverse geometries \cite{Rothemund2006,Ke2009SquareOrigami,Tian2015,Ketterer2018}. Each origami structure contains programmable ssDNA binding sites that can be selectively functionalized with molecular spin labels or other qubit candidates. (b) Schematic of the hybrid platform, where spin-bearing molecules patterned on DNA origami are positioned in proximity to shallow NV centers in diamond, providing optical addressability at the single spin level. (c) Site-specific immobilization of single molecules enables nanoscale magnetic resonance imaging with NV sensors. (d) Arrays of aptamers attached via programmable binding sites allow high-throughput proteomic assays, with nearby NV sensors reporting on binding events. (e) Ordered arrays of molecular spins can be engineered to generate entangled states (such as spin-squeezed states), enhancing quantum sensing sensitivity beyond the standard quantum limit. (f) DNA-templated emitter arrays with subwavelengnth spacing ($a \ll \lambda$) open access to collective photonic phenomena, including superradiance, subradiance, and topologically protected edge modes. 
}
\end{figure}

DNA origami (Fig.~\ref{fig:1}a) presents an emerging and largely unexplored approach to the challenge of positioning atomic-scale qubits with nanoscale precision. 
Originally developed as a structural breadboard for organizing matter at the nanoscale \cite{Rothemund2006,Douglas2009,Dietz2009}, DNA origami enables bottom-up, solution-phase assembly of custom architectures (in 1, 2 or 3 dimensions) with nanometer-scale spatial control. These structures are formed by folding a long single-stranded DNA (ssDNA) ``scaffold'' into a prescribed shape using short synthetic ssDNA ``staples'' to secure the proximity of distant scaffold domains through base pairing. 
While this technique has been extensively developed for applications in nanophotonics, molecular electronics, and biosensing \cite{Knappe2023,Zhan2023,Gopinath2016,Gopinath2021,Martynenko2023}, its integration with coherent spin-based quantum technologies remains minimally explored. 
Yet DNA origami offers a uniquely compelling platform for organizing spin qubits (Fig. \ref{fig:1}b), supporting large-scale, highly tunable, and massively parallel patterning of molecular systems under mild and chemically flexible conditions. 
Further, its compatibility with surface patterning \cite{Gopinath2014,Gopinath2016,Shetty2021} raises the prospect of a hybrid quantum device in which single, DNA-patterned molecular spins could be initialized, manipulated, and read out via coupling to near-surface optically active defect-based spins, such as the well-established NV center in diamond.

Fig. \ref{fig:1}c-e illustrates several quantum technologies enabled by the nanoscale positioning capabilities of DNA origami, ranging from quantum sensing to the engineering of many-body quantum dynamics. In the context of sensing (Figs. \ref{fig:1}c,d), DNA origami enables two key applications: single-molecule magnetic resonance imaging \cite{Hollenberg2014, Lovchinsky2016, DegenReview2017} and high-throughput proteomics, in which structure-switching of spin-labeled aptamers upon binding to target proteins \cite{Dunn2017_review_Aptamer} is detected by a sub-surface NV magnetometer with single-spin detection capabilities \cite{lu_magnetically_2023}. 
These holy-grail sensing tasks are actively pursued today, but progress has been stymied by the lack of control over target density, surface proximity, and binding chemistry \cite{AwschalomReview2021, JanitzReview2022}. A DNA origami - diamond NV hybrid platform provides site-specific immobilization of molecular targets with nanometer-scale proximity to NV sensors. 

The ability to form regular nanoscale, spin-qubit arrays using DNA origami suggests the complementary perspective (Fig.~\ref{fig:1}e-f) of viewing our platform as a quantum simulator capable of exploring strongly-correlated phenomena. In atomic systems, exquisite spatial control over single atoms has enabled transformative advances in quantum sensing and many-body simulation, including the generation of spin-squeezed states that surpass the standard quantum limit (SQL) in magnetometry ~\cite{bornet_squeezing_2023}. Recent theoretical and experimental work has highlighted the critical role of dimensionality and spatial ordering in realizing dipolar-driven spin-squeezing \cite{Block2024,kobrin2024universalprotocolquantumenhancedsensing, wu_squeezing_2025, gao_squeezing_2025}, posing a challenge for translating these techniques to solid-state spin implementations. DNA origami offers a unique approach to this challenge. In a related direction (Fig.~\ref{fig:1}f), theoretical proposals and early experiments have predicted and demonstrated emergent photonic behavior in sub-wavelength-spaced ordered arrays of dipolar-interacting single-photon emitters, such as superradiance, subradiance, and topologically protected edge modes \cite{Perczel2017, DaqingWang2019, Rui2020, Srakaew2023, YoungshinKim2024}. However, a key limitation of atomic platforms is that emitter spacing is constrained by the wavelength of the trapping light, typically hundreds of nanometers. DNA origami patterned with chemically functionalized emitters provides a complementary approach, enabling emitter spacing well below this limit. These capabilities position DNA origami as a powerful and versatile tool for deterministic positioning of spins and emitters, with significant untapped potential for advancing quantum technologies.

This work demonstrates the ability of DNA origami to template molecular spins and read them out with single spin sensitivity, critically enabled by the compatibility of DNA-functionalized diamond surfaces with coherent NV center quantum sensors. 
We use self-assembled DNA origami to place Gd$^{3+}$ molecular spins on a diamond surface with control over their number and position via the programmable nature of the origami binding sites. We use shallow NV center ensembles residing 5 nm below the origami-functionalized diamond surface as local, optically detectable probes of Gd$^{3+}$, a large electronic spin that induces $T_1$ relaxation of the NV centers.  
We first confirm that bare origami (without Gd$^{3+}$) neither adversely affects NV coherence nor modifies the local electronic spin environment, and hence is an excellent surface functionalization approach. We then detect origami-templated Gd$^{3+}$ via its induced relaxation of the NV center and find a linear dependence of this signal on Gd$^{3+}$ spin density, thus showing the spin patterning capability of DNA. 
Importantly, even under DNA origami functionalization, our shallow NV centers remain sensitive to less than 1 Gd$^{3+}$ per NV sensing volume. Lastly, we discuss and quantitatively analyze how two quantum sensing applications -- high-throughput proteomics and entanglement-based sub-SQL sensing -- could be enabled by the integration of DNA origami with solid-state and molecular qubits.

Our experimental platform consists of 2-dimensional DNA origami deposited and dried on a single-crystal diamond surface that contains a shallow NV ensemble (see Methods and SI). The DNA origami (Fig.~\ref{fig:2}a) unit is an equilateral triangular shape of $\sim$130 nm side length with 204 programmable binding sites on one surface \cite{Gopinath2021}, each of which can be functionalized with 0 to 4 Gd$^{3+}$ spins. Control experiments were performed with unlabeled origami. After depositing DNA origami on the diamond surface, atomic force microscopy (AFM) was used to assess the origami coverage at the same locations where confocal measurements were performed (colocalization is achieved with diamond surface features as discussed in the SI). 
We use a home-built, room-temperature confocal microscope with 532-nm laser excitation for optically addressing NV centers \cite{BluvsteinSurfaceDriving}.

\section{Integrating DNA origami with diamond qubits}

\begin{figure}[b]
\centering
\includegraphics[width=86mm]{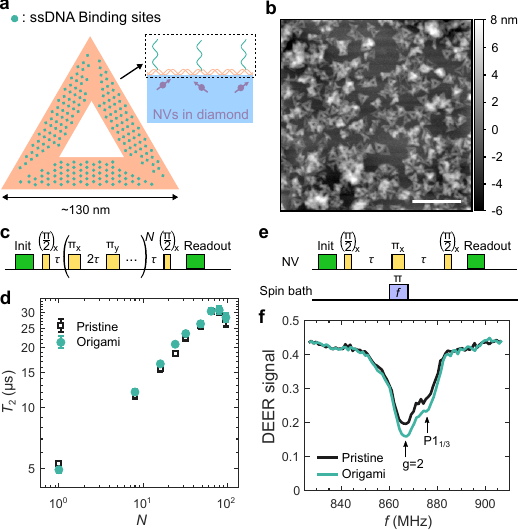}
\caption{\label{fig:2} \textbf{DNA origami preserves the coherence and spin environment of shallow NV centers.}
(a) Schematic of triangular DNA origami (orange) with 204 programmable ssDNA binding sites (top down view). Inset: Side view schematic showing a zoom in of origami deposited on the surface of diamond with a shallow NV center ensemble. (b) AFM image of the diamond surface after origami deposition and drying, with $\sim$62\% coverage. Scale bar: 500 nm. (c) Pulse sequence for coherence measurements: $N = 1$ corresponds to Hahn echo, while $N \geq 8$ implements XY8-$k$ ($N=8k$) dynamical decoupling. (d) Comparison of NV $T_2$ as a function of $N$ for NVs under a bare diamond surface and under unlabeled origami, showing no measurable degradation under unlabeled origami. (e) Pulse sequence for DEER spectroscopy, where RF frequency $f$ is swept to selectively recouple bath spins. (f) DEER spectra of pristine and origami-coated surfaces, both exhibiting the characteristic $g=2$ surface spin signal and $P1$ bulk features with no new resonances.}
\end{figure}

To assess the suitability of DNA origami as a surface functionalization strategy for NV-based quantum sensing, one must first establish that its presence does not degrade the spin, charge, and optical properties of shallow NVs, nor does it modify the NVs' local spin environment. Near-surface NVs are highly sensitive to changes in their local environment \cite{Rosskopf2014,Sangtawesin2019}, and many existing surface functionalization methods are known to perturb their coherence times and charge stability \cite{Mouzhe2022, rodgers_diamond_2024}. Here, we perform several measurements to evaluate the effects of DNA origami on shallow NV centers. Fig.~\ref{fig:2}c-f focus on two key measurements: coherence decay under dynamical decoupling, which probes broadband magnetic noise, and double electron-electron resonance (DEER), which probes the density and spectral properties of local paramagnetic spins. 

We prepare the substrate by depositing unlabeled DNA origami (without Gd$^{3+}$) onto an oxygen-terminated diamond surface containing a $\sim$5 nm deep NV ensemble (Fig.~\ref{fig:2}a). AFM imaging confirms uniform coverage of triangular origami structures with an average surface coverage of 62\% (Fig.~\ref{fig:2}b), and co-localization with confocal features ensures that NV measurements are performed on NVs directly beneath the origami-decorated surface. We first probe NV coherence by measuring $T_2$ as a function of the number of applied $\pi$ pulses, $N$. The case $N=1$ corresponds to a Hahn echo, while for $N \geq 8$ we employ XY8-$k$ dynamical decoupling with $N=8k$. As shown in Fig.~\ref{fig:2}d, we observe no significant change of the coherence time under DNA origami for either Hahn echo or XY8-$k$. Moreover, the functional form of the decay curves remains unchanged. These observations confirm that DNA origami does not introduce additional broadband magnetic noise near the surface. We next characterize the local electronic spin environment using DEER spectroscopy (Fig.~\ref{fig:2}f). The spectrum exhibits the ubiquitous g = 2 resonance from native surface electronic spins \cite{Mamin2012}, along with features from substitutional nitrogen (P1) centers in the diamond. Upon origami deposition, the amplitude of the g = 2 peak increases modestly by 21.1$\pm$5.8\% while the change in FWHM from 13.92$\pm$0.48 MHz to 14.54$\pm$0.51 MHz is within the measurement uncertainty. Similarly, the P1 peak shows no significant change in amplitude (16.3$\pm$20.5\%) or FWHM (6.90$\pm$0.70 MHz to 6.93$\pm$0.71 MHz). Importantly, no additional spectral features or broadening appear following functionalization, indicating that DNA origami does not introduce a new population of deleterious electronic spins or surface disorder. We observe minimal changes in other NV properties, including photoluminescence (PL) intensity and Rabi contrast (see SI). 

Together, these results demonstrate that DNA origami provides a surface functionalization architecture that preserves shallow NV coherence and charge state stability, maintains the native surface spin environment, and leaves NV photophysics intact. This robustness likely originates from the compatibility between DNA origami and the oxygenated surfaces required to stabilize shallow NV centers. In the presence of Mg$^{2+}$ ions in the buffer, the negatively charged origami structures adhere naturally to the negatively charged surface via electrostatic bridging \cite{Gopinath2016}, thereby preserving the near-surface band behavior.

\section{DNA-driven control of molecular spin spacing}

\begin{figure}[b]
\includegraphics[width=86mm]{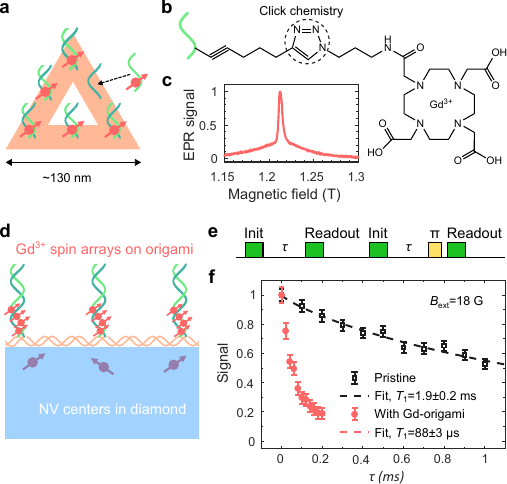}
\centering

\caption{\label{fig:3} \textbf{Detection of DNA-origami-templated Gd$^{3+}$ with NV relaxometry.} (a) Triangular DNA origami functionalized with Gd-DOTA (red) via hybridization of Gd-labeled ssDNA to programmable binding sites. (b) Gd-DOTA structure and conjugation to ssDNA (light green) by click chemistry. (c) Continuous-wave EPR spectrum of Gd-labeled origami in solution at Q-band (33.81 GHz), yielding $g = 1.9923(3)$, the expected value for Gd-DOTA and distinct from the bare electron g-factor. (d) Schematic of Gd$^{3+}$ arrays positioned on DNA origami above shallow NV centers in diamond. Shown are 4 Gd$^{3+}$ per site. (e) Pulse sequence for differential $T_1$ relaxometry, showing microwave $\pi$-pulses (yellow) tuned to the $|m_{s}=0\rangle \leftrightarrow |m_{s}=-1\rangle$ transition of an NV aligned to $B_\text{ext} = 18$G, and optical pulses (green) for initialization and readout. (f) NV $T_1$ decay before and after deposition of Gd-labeled origami, showing a reduction from $1.9 \pm 0.2$ ms with stretch exponent $n = 0.83(7)$ (pristine) to $88 \pm 3$ $\mu$s with $n = 0.75(4)$ (4 Gd$^{3+}$ per binding site on each of the 204 binding sites).}
\end{figure}

\begin{figure}[b]
\centering
\includegraphics[width=86mm]{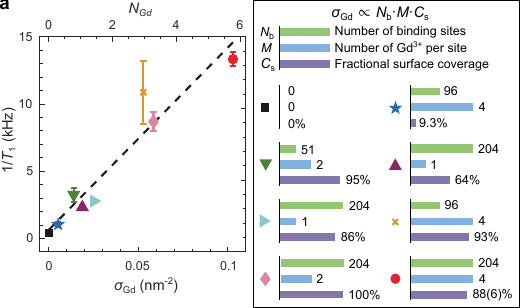}
\caption{\label{fig:4} \textbf{Programmable control of Gd$^{3+}$ spin density via DNA origami.} (a) NV relaxation rate $1/T_1$ measured across eight distinct DNA origami designs with varying numbers of binding sites ($N_B$), Gd$^{3+}$ per site ($M$), and fractional surface coverage ($C_S$), with values indicated on the right panel. $1/T_1$ data increases linearly with the engineered Gd$^{3+}$ surface density $\sigma_\text{gd} \propto N_b\cdot M \cdot C_S$ in agreement with numerical simulations (dashed line). The upper axis indicates the effective number of Gd$^{3+}$ spins $N_{\text{Gd}}$ within the sensing volume of a 4.5 nm deep NV (see SI). Error bars represent standard deviation across multiple diamond locations. 
}
\end{figure}

Having established the viability of a DNA origami-based surface functionalization architecture for shallow NV centers, we next demonstrate that this platform enables controlled interfacing between NVs and patterned molecular spins. Our goal is to leverage the programmability of DNA origami to chemically modulate the density of spin-active molecules on the diamond surface, and to verify this capability via NV-based readout. Each of the 204 programmable ssDNA binding sites on one side of the triangular DNA origami can be selectively hybridized with complementary spin-label-functionalized ssDNA. We attach Gd-DOTA spin labels to the binding strands using a standard click chemistry procedure (Methods and Fig.~\ref{fig:3}b) resulting in a designable 2D array of Gd$^{3+}$ ions on the origami breadboard (Fig.~\ref{fig:3}a).
By varying the number of occupied binding sites ($N_b$), we determine which positions on the origami carry spins, thereby controlling their relative spacing. Independently, by varying the number of Gd$^{3+}$ ions per site ($M$), we adjust the spin multiplicity at each occupied position, thereby tuning the overall density without changing spacing. This dual control lets us prescribe the total spin content per origami while also specifying its geometry. Continuous-wave EPR spectroscopy (Fig.~\ref{fig:3}c) confirms successful labeling of the origami with Gd$^{3+}$, with a clear resonance at 1.2125(2) T (g = 1.9923(3)), consistent with isolated Gd-DOTA \cite{Caravan1999,Merbach2000, Merbach2001, Clayton2018}.

To validate this design locally via a measurable NV response, we deposit the hybridized origami on the diamond surface and perform NV $T_1$ relaxometry. Gd$^{3+}$ ions, with large electronic spin of S = 7/2 and fast gigahertz-scale dynamics, generate broadband magnetic noise at the NV zero-field splitting (2.87 GHz), which drives spin relaxation. 
Each NV center detects Gd$^{3+}$ ions primarily within a sensing area of $\sim$ 4 nm radius on the diamond surface (the region contributing 70\% of the relaxation signal) and our confocal measurement probes an ensemble of NV centers in a spot with $\sim$ 300 nm radius. The relaxation rate of an NV center in the presence of a proximal Gd$^{3+}$ spin, which can be modeled with a Lorentzian noise spectrum \cite{Slichter1978,Schoelkopf2003,Tetienne2013PRB}, is given by
\begin{equation}\label{T1_eq} \frac{1}{T_{1}}=3\Omega'+\left(\frac{\mu_0\hbar\gamma_\text{nv}\gamma_{\text{Gd}}}{4\pi r^3}\right)^2\frac{(S^2+S)(2+3\sin^2{\alpha})\tau_\text{c}}{1+\omega_0^2\tau^2_\text{c}}, \end{equation}
where $\Omega'$ is the intrinsic relaxation rate of the NV ground state manifold ($\ket{m_s=0} \leftrightarrow \ket{m_s=\pm 1}$), $\gamma_{\text{nv}}$ and $\gamma_{\text{Gd}}$ are the gyromagnetic ratios, $\mu_0$ is the Bohr magneton, $\alpha$ is the angle between the NV quantization axis and the NV-Gd direction, $\omega_0$ is the NV transition frequency, $\tau_c$ is the correlation time of Gd$^{3+}$, and $r$ is the NV–Gd separation. We perform a differential $T_1$ measurement, which uses a common-mode rejection technique to disambiguate the spin-induced changes in relaxation from charge-related mechanisms \cite{myers2017, BluvsteinChargeState}, and fit the $T_1$ spin decay to $\exp[-({\tau/T_1})^n]$. Without Gd$^{3+}$, we observe an intrinsic $T_1$ of 1.9(2) ms with a stretched exponential decay ($n = 0.83(7)$, where $n<1$ is consistent with ensemble averaging over a distribution of decay rates \cite{Choi2017}). Indeed, after depositing Gd-labeled origami, the $T_1$ at the same location drops sharply to 88(2) $\mu$s (Fig.~\ref{fig:3}f), confirming the presence of Gd$^{3+}$.

We next evaluate the degree of positional control by performing experiments across a range of origami configurations, systematically varying $N_b$, $M$, and the fractional surface coverage $C_s$ (as defined and quantified via AFM in Methods). Because each NV is sensitive to the number of Gd$^{3+}$ in its sensing area, a natural quantity to invoke is the average surface density of Gd$^{3+}$, defined as
\begin{equation}
    \sigma_{\text{Gd}}=\frac{N_\text{b}\cdot M\cdot C_\text{s}}{A},
\end{equation}
where $A$ is the projected area of a single triangle. Eight total configurations of $N_b$, $M$, $C_s$ were tested, with $T_1$ measured at seven positions per configuration to extract the average relaxation rate $1/T_1$ (Fig.~\ref{fig:4}a).
Before each measurement under new DNA origami deposition conditions, we reset the diamond surface by cleaning in water and boiling piranha (H$_2$SO$_4$:H$_2$O$_2$) solution.
We observe a strongly linear correlation between the engineered Gd$^{3+}$ density $\sigma_{\text{Gd}}$ and the measured relaxation rate, with a Pearson correlation coefficient of $r_P = 0.958$. We further investigate this linear dependence by performing Monte Carlo simulations using  Eqn.~\ref{T1_eq}, incorporating NV depth distributions based on Stopping and Range of Ions in Matter (SRIM) simulations and assuming Gd$^{3+}$ locations on origami binding sites. The simulation (dashed line in Fig.~\ref{fig:4}a) reproduces the observed linear scaling of $1/T_1$ with $\sigma_{\text{Gd}}$, where the slope is solely determined by $\tau_c$; the best fit gives $\tau_c=0.18\pm0.03\text{ ns}$, consistent with values in the literature \cite{Rast2001,Li_Gd_review_2019}. The observed linear relationship confirms that the NV relaxation signal quantitatively reflects the engineered spin density, while the underlying control over spin spacing that is intrinsic to the design is latent in these measurements. 

Several key features emerge from this experiment. First, the ability to tune the density and position of patterned spins by independently controlling $N_b$, $M$, and $C_s$ demonstrates a degree of programmability not previously achieved in NV-based surface sensing platforms. In particular, the fact that $N_b$ determines which sites are labeled means that, beyond density control, this platform offers deterministic placement of spins at prescribed separations—an essential ingredient for engineering dipolar interactions and collective effects. Moreover, by converting the spin density into  the effective number of Gd$^{3+}$ spins within each NV’s sensing area ($N_{Gd}$ as shown on the top axis of Fig.~\ref{fig:3}a), we find that NVs are sensitive to fewer than one spin per sensing area. This sensitivity to single spins is enabled by two core aspects of the NV-origami architecture: NVs maintain their intrinsic $T_1$ and charge state properties under unlabeled origami, and the low-profile geometry of the origami only adds 1-2 nm of separation between the NVs and target spins. This minimal spacing is particularly critical for $T_1$ relaxometry, where the signal strength from a single spin falls off as $1/r^6$.

\section{Future applications enabled by DNA origami}

\begin{figure}
\includegraphics[width=86mm]{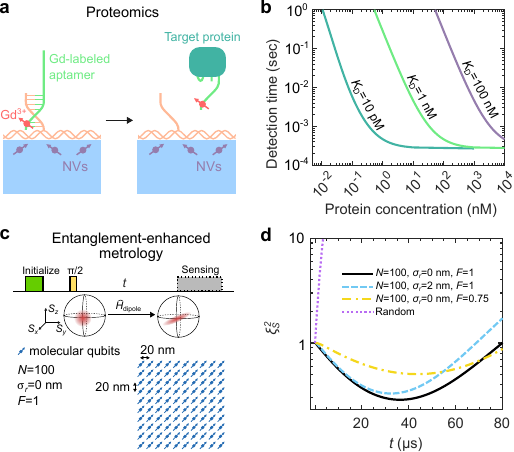}
\centering
\caption{\label{fig:5} \textbf{Versatile quantum sensing enabled by DNA origami patterning.} (a) Schematic of a label-free proteomics assay based on $T_1$ relaxometry of shallow NV centers. DNA origami on the diamond surface immobilizes Gd$^{3+}$-labeled aptamers. Target binding causes aptamers to dissociate, displacing the spin label and increasing $T_1$. (b) Simulated detection time versus protein concentration for three binding affinities, illustrating the potential for rapid, parallelized sensing. (c) Schematic of a dipolar-interacting spin array for entanglement-enhanced metrology. Molecular qubits are patterned into a square lattice of $N = 100$ spins, with positional uncertainty $\sigma_r$ and binding site filling rate $F$. Quenching a coherent spin state along $+x$ under its native dipolar Hamiltonian prepares the squeezed state for sensing. (d) The simulated squeezing parameter as a function of evolution time $t$. Ideal spatial ordering produces squeezing ($\xi_S^2 < 1$), while moderate disorder or partial filling reduces but maintains squeezing. Random placement yields no squeezing.}
\end{figure}

The DNA origami platform enables versatile quantum sensing applications by deterministically placing functional molecules with nanometer precision above shallow NV centers. We illustrate two key directions in Fig.~\ref{fig:5}: high throughput proteomics via massively parallel detection of biomolecular binding events using aptamers, and entanglement-assisted quantum metrology via the construction of programmable arrays of dipolar-coupled spins.

Reliable proteomics measurements, essential for early disease detection and personalized medicine, require highly specific, parallel detection of large panels of molecular targets using minimal sample volumes. Recent advances in aptamer chemistry, including the development of SOMAmers with nanomolar to picomolar dissociation constants, enable selective molecular recognition across a wide range of proteins \cite{Dunn2017_review_Aptamer,lu_magnetically_2023}. These short, synthetic DNA or RNA sequences bind specific protein targets and undergo conformational changes upon binding. In our proposed scheme inspired by previous works \cite{lu_magnetically_2023,chi-duran_multiplexed_2025}, aptamers labeled with Gd$^{3+}$ spin tags are initially hybridized to the origami structure positioned above a shallow NV ensemble. Upon binding to their targets the aptamers dissociate from the origami, displacing the Gd$^{3+}$ label from the NV vicinity and increasing the NV's $T_1$. 
This label-free detection scheme allows for rapid multiplexing via spatial patterning of diverse aptamer sequences across the diamond. We note that multiplexed functionalized diamond microarrays have recently been demonstrated, with NV relaxometry establishing a direct link between biomolecular target detection and quantum readout \cite{chi-duran_multiplexed_2025}. In Fig.~\ref{fig:5}b, we simulate the minimal detection time for various protein affinities under realistic assumptions: a $1\ \mu\text{L}$ sample volume covering a 4-mm$^2$ diamond surface functionalized with approximately $1.25 \times 10^{-17}\ \text{mol}$ of aptamers on a $300\ \mu\text{m}^2$ sensing area, achievable using triangular origami with a binding site density of approximately $0.026\ \text{nm}^{-2}$. We assume an NV ensemble readout of $50$ counts per shot over one sensing area, and photon shot-noise–limited detection with $5\ \text{ms}$ overhead per shot. Full modeling assumptions and derivations are provided in the SI. The short detection timescales combined with our platform's single molecule detection capabilities establish the NV-origami quantum sensor platform as a new tool for biosensing with minimal sample volume required.

We next discuss prospects for entanglement-enhanced metrology (Fig.~\ref{fig:5}c-d). An ensemble of spins positionally-ordered with DNA origami exhibiting strong spin-spin dipolar interactions is a rich playground for studying many-body quantum dynamics. Recently, it has been proposed~\cite{Block2024} that metrologically useful many-body entangled states can be realized in such strongly interacting spin arrays, and spin squeezing~\cite{kitagawaSqueezedSpinStates1993} has been experimentally demonstrated in both dipolar atomic~\cite{bornet_squeezing_2023} and solid-state systems~\cite{wu_squeezing_2025}. However, in the solid-state case, the amount of achievable squeezing was significantly constrained by positional disorder. By contrast, DNA origami enables precise spatial control over spin placement, mitigating disorder-induced limitations and enabling geometries optimized for entanglement generation. In addition, its nanometer-scale proximity to a surface provides a natural route for integrating such squeezed states with sensing targets, offering a path toward quantum-enhanced metrology in realistic environments.

We numerically explore this prospect for DNA-templated molecular spins by simulating a square lattice of spin-$\tfrac{1}{2}$ molecular qubits with quantization axis perpendicular to the 2D plane and a lattice constant of $20\ \text{nm}$—well within the resolution achievable by origami patterning—and computing the system’s time evolution under the magnetic dipole-dipole interaction (Fig.~\ref{fig:5}c). 
We envision that the molecular spins are initialized in a coherent spin state polarized along the $+x$ direction of the Bloch sphere, and that the system subsequently evolves under the dipolar Hamiltonian:
\begin{equation}
\hat{H}_{\text{dipole}} = -\sum_{i \neq j} \frac{J}{r_{i,j}^3} \left( \hat{S}^x_i \hat{S}^x_j + \hat{S}^y_i \hat{S}^y_j - 2 \hat{S}^z_i \hat{S}^z_j \right),
\end{equation}
where $J = (2\pi)\times 52\ \text{MHz}\cdot\text{nm}^3$ is the dipolar coupling constant, $r_{i,j}$ is the distance between spins $i$ and $j$, and $\hat{S}_i$ are the spin-$\tfrac{1}{2}$ operators. We simulate the dynamics using the discrete truncated Wigner approximation (DTWA) \cite{Schachenmayer2015}, which captures mean-field correlations and reliably predicts the onset of spin squeezing in dipolar systems \cite{Perlin2020,Block2024}.
To characterize squeezing, we compute the Wineland squeezing parameter
\begin{equation}
\xi_S^2 = N  \frac{\min_\theta \text{Var}(S_\theta)}{S_x^2},
\end{equation}
where $N$ is the total number of spins, and $S_\theta$ is the collective spin operator along an axis orthogonal to the mean spin direction ($+x$). As shown in Fig.~\ref{fig:5}d, $\xi_S^2$ drops below unity in the ideal lattice configuration, indicating a sensitivity beyond the SQL. Moderate imperfections, including positional disorder $\sigma_r = 2\ \text{nm}$ and partial occupancy $F = 0.75$ diminish but do not eliminate the spin squeezing. In contrast, a control case with spins randomly placed at the same areal density exhibits no squeezing. These results highlight the unique capacity of DNA origami to impose spatial order at the nanoscale, enabling entanglement-enhanced quantum metrology via dipolar-driven, many-body interactions.

\section{Discussion}

We propose a novel experimental platform that combines molecular qubits, DNA origami, and solid-state quantum sensors. DNA origami uniquely acts as a breadboard for the deterministic patterning of single molecules on the nanometer scale. We first verify that the deposition of DNA origami on the diamond surface does not degrade the NV properties. Furthermore, we observe a strongly linear correlation between the NV relaxation time and the controlled density of Gd$^{3+}$ ions that we patterned on the diamond surface using DNA origami. This novel experiment platform is promising for several future directions, including studying many-body physics in strongly interacting systems and functionalizing diamond surfaces for quantum sensing applications.
Looking forward to molecular biosensing of relevant protein targets, next steps will involve combining our hybrid NV-DNA origami platform with Gd-labeled aptamers and target proteins. 

For studies and applications of coherent many-body dynamics, future work will need to interface DNA origami with coherent molecular qubits. Our approach is agnostic to spin species, so any moiety that can be functionalized to ssDNA becomes accessible. Coherent candidates such as trityl radicals, endohedral fullerenes, or even Gd$^{3+}$ itself (acquiring long coherence times at cryogenic temperatures and high magnetic fields \cite{wilson_gadolinium_2023}) could be incorporated with no change to the fabrication workflow. We envision patterned spins being initialized and read out via coupling to sub-surface NV centers \cite{pinto_readout_2020} or via native optical transitions \cite{Bayliss2020, feder_fluorescent-protein_2025}. Future experiments will measure the coherent dynamics of these patterned spin ensembles as a function of spin spacing and geometry. 

The convergence of deterministic molecular placement, spin coherence, and quantum readout opens a new frontier for quantum sensing. Critically, the same features that make DNA origami attractive for multiplexed proteomics (site-specific functionalization, high-density encoding, and nanoscale addressability) could also enable the generation and deployment of entangled spin ensembles in close proximity to target molecules, cells, or condensed matter systems, offering a route to perform quantum-enhanced sensing directly at the interface with physical systems of interest. DNA origami, long celebrated for its programmability in structural biology, now stands poised to reshape the material landscape of quantum science.

\section{Methods}
\setcounter{secnumdepth}{2}
\subsection{Diamond sample fabrication and characterization}
Diamond homoepitaxial growth was performed via plasma-enhanced chemical vapor deposition (PECVD) using a SEKI SDS6300 reactor on a [100] oriented electronic grade diamond substrate (Element Six Ltd.). Prior to growth, the substrate was fine-polished by Syntek Ltd. to a surface roughness of 200-300 pm, followed by a 500 nm etch to relieve polishing-induced strain. The growth conditions consisted of a 750 W plasma containing 0.1$\%$ $^{12}$CH$_{4}$ in 400 sccm H$_2$ flow held at 25 torr and $\sim$800 $^{\circ}$C. A $\sim$120 nm-thick isotopically purified (99.98$\%$ $^{12}$C) epilayer was grown. After growth, to create a dense two-dimensional shallow NV ensemble, the sample was implanted with $^{14}$N (2.5 keV and 1e13 cm$^{-2}$ dose) by INNOVion and annealed in an Ar/H$_{2}$ atmosphere at 850$^{\circ}$C for 2 hours. After this first round of high-temperature annealing, we found the NV $T_2$ of a specific location was 1.8 \textmu s. Then, the sample is annealed again at 1200$^{\circ}$C for 2 hours. After annealing, the sample is cleaned in tri-acid ($\text{HClO}_4$:$\text{H}_2\text{NO}_3$:$\text{H}_2\text{SO}_4$ 1:1:1) for 1 hour at 150$^{\circ}$C, then annealed in air at 450$^{\circ}$C for 4 hours. We observed the NV $T_2$ from the same location increased to 2.2 \textmu s. Then, to further augment the NV density, we performed local electron irradiation with a transmission electron microscope (Talos F200X G2 TEM) over several areas of this sample with different dosages at 200 keV\cite{Mainwood1994,Deak2014,Eichhorn2019,Hughes2023}. After irradiation, the sample is then annealed in an Ar/H$_{2}$ atmosphere at 1200$^{\circ}$C for another 2 hours, cleaned again in tri-acid for 1 hour at 150$^{\circ}$C, annealed in air at 450$^{\circ}$C for 4 hours, and cleaned in boiling piranha solution ($\text{H}_2\text{SO}_4$:$\text{H}_2\text{O}_2$ 3:1) for 20 min. The final NV $T_2$ from the same location increased to 3.4 \textmu s.  The diamond was always reset by boiling it in piranha solution for 20 min before depositing a new DNA origami sample. We also performed another tri-acid cleaning after about 15 rounds of origami deposition and measurement to fully recover the NV charge state stability.

\subsection{Preparation of Gd$^{3+}$ on DNA origami}
In this work, we selected a widely used triangular origami design due to its ease of identification under atomic force microscopy (AFM), structural rigidity, and low aggregation tendency \cite{Rothemund2006, Gopinath2016}. The details of our design are most similar to those in \cite{Gopinath2016}. Gd$^{3+}$ was patterned on the origami via hybridization between a pair of complementary ssDNA sequences 29-bases long.
One sequence - the binding site - was appended to the 5' end of all, or any desired subset, of the origami staples, so that it emerged from the origami surface at specified sites. This sequence (5'-AATGCTGATGCAATGTGCGCAAATAAAAA-3') was designed to minimize the likelihood of origami hybridizing to one another. The complementary sequence (5'-TTTTTATTTGCGCACATTGCATCAGCATT-3') - the label - was modified at its 5' end (as elaborated below) so that the Gd$^{3+}$ were held as close as possible to the origami surface upon hybridization.
\setcounter{secnumdepth}{3}
\subsubsection{Triangle origami self-assembly and purification} 
Staple strands (Integrated DNA Technologies) and the scaffold strand (single-stranded M13mp18, Bayou Biolabs, P-107) were mixed together to target concentrations of 100 nM (each staple) and 40 nM, respectively (staple:scaffold 2.5:1), in ‘Tris/Mg$^{2+}$ buffer’ (10 mM Tris Base with 12.5 mM magnesium chloride, adjusted to pH 8.35-8.40 with HCl). 
50 \textmu L volumes of this staple/scaffold mixture were placed in 0.5 mL DNA LoBind tubes (Eppendorf), heated to 90°C for 5~min and annealed from 90°C to 20°C at -0.2°C/min in a PCR machine. Excess staples were removed by centrifugation through 100 kD molecular weight cut-off spin filters (Amicon Ultra-0.5 mL). All centrifugation steps were performed at 2,000g for 8.5~min at room temperature. The best yields (35\%–50\% with no staples visible by agarose gel) were achieved by first wetting the filter (adding 500 \textmu L Tris/Mg$^{2+}$ buffer and centrifuging until the retained volume was $\sim$50 \textmu L and discarding the filtrate). The pre-wet filter was loaded by adding 50 \textmu L of unpurified origami and 400 \textmu L Tris/Mg$^{2+}$ buffer, centrifuging and removing the filtrate. The loaded origami was then washed twice with 400 \textmu L Tris/Mg$^{2+}$ buffer and, finally, recovered by inverting the filter into a new 0.5 mL DNA LoBind tube and centrifuging. Throughout this process, the amount of DNA in the filtrate and in the final the origami stock solution was monitored in terms of A$_{260}$ by microvolume spectrometry (NanoDrop 2000, Thermo Scientific). 

\subsubsection{Gd$^{3+}$-label strand preparation and patterning}
Label strands (Integrated DNA Technologies) were purchased with between 0 and 4 of the thymine bases at the 5' end replaced by octadiynyl dU and suspended in Milli-Q water at $\sim$0.5 mM. 
Gd-Azido-mono-amide-DOTA (hereafter referred to as Gd-DOTA)(Macrocyclics, Inc.) was suspended in Milli-Q water at $\sim$100 mM.  
Click reagent was prepared by dissolving BTTAA (Bioconjugate Technologies, LLC.) and CuCl$_2$·2H$_2$O (Sigma-Aldrich) in Milli-Q water to yield BTTAA:CuCl$_2$ 1:1 solution. 
Gd-DOTA was covalently `clicked' to the octadiynyl dU on the label strands as follows: 
First, 200 \textmu L of 0.5 mM label strand with $M$ modified thymine bases (ratio=1) 
is mixed with $5M$ \textmu L of 100 mM Gd-DOTA (ratio=$5M$) in a 0.5 mL LoBind tube (Eppendorf). Then, 5 \textmu L of 10 mM BTTAA:CuCl$_2$ 1:1 solution (ratio=4), and 20 \textmu L of 100 mM fresh Sodium Ascorbate (MilliporeSigma) in Milli-Q water (ratio=20) were added. Next, a clean syringe was used to bubble N$_2$ through the solution for 5 min to purge dissolved O$_2$. Successful purging was assessed from the transparent or light yellow (instead of blue or green) color of the solution and the absence of precipitate. Finally, the solution was incubated for at least 10 hours at room temperature. 

The Gd-labeled ssDNA was purified into Tris/Mg$^{2+}$ buffer by centrifugation through a 3 kD molecular weight cut-off spin filter (Amicon Ultra-0.5 mL) using the same procedure as for origami purification described above, except that all centrifugation was done at 14,000g for 15~min. The purity of the resulting Gd-labeled ssDNA was verified by liquid chromatography-mass spectroscopy (LC-MS, Novatia, LLC). 

Purified origami and purified Gd-labeled strands were mixed in a 2:3 ratio and left to hybridize at room temperature for at least 12 hours. Unhybridized label strands were removed by centrifugation through a 100 kDa molecular weight cut-off spin filters (Amicon Ultra-0.5 mL), using the same procedure as for origami purification described above. 

\subsection{Labeled origami deposition on diamond}
The diamond was positioned at the center of a clean polypropylene wafer carrier, with its NV side facing up. A 2 \textmu L drop of deposition buffer (100 mM Tris Base with 150 mM magnesium chloride, adjusted to pH 8.9 with HCl) was pipetted onto its surface.
Then, 2 \textmu L of purified labeled origami was added and mixed into the drop by gently pipetting up and down for 1 min. Labeled origami was then left to settle onto the diamond surface by incubating at room temperature for 15 min. The diamond was then immersed in 150 mM Tris/Mg$^{2+}$ pH=8.90 buffer for 5 sec, transferred into 75\% isopropyl alcohol (IPA) in Milli-Q water (v/v\%) for 5 sec, and finally into 100\% IPA for 1 min, then left to air dry on a lint-free cloth (Kimtech Kimwipes).

When deposition was not successful, NV $T_1$ did not drop below 1 ms. 
The fractional surface coverage of origami on the diamond, $C_\text{s}$, was assessed by atomic force microscopy (AFM). The triangular hole in the center of the origami design (Fig.~\ref{fig:2}b) limits the maximal coverage to 86\% of the diamond surface.  
We therefore calculated the fractional surface coverage as
\[
C_\text{s} = A_\text{dna}/(0.86\times A_\text{scan})
\]
where $A_\text{dna}$ is the area that is at least 1 nm above the diamond surface in the AFM image and  $A_\text{scan}$ is the AFM scan size. The extent of surface coverage increased with the concentration of origami solution placed on top of the diamond. Origami concentrations $\geq$4 nM routinely yielded $C_\text{s} = 88(6)\%$ was estimated from the AFM images. 

\section{Acknowledgments}

We thank Dr. Blake Wilson (UCSB), Dr. Mark Sherwin (UCSB), Dr. Chung-Ta Han (Northwestern), and Dr. Songi Han (Northwestern) for insightful discussion and help with EPR measurements.
This work was supported by the Army Research Office through the MURI program Grant No.~W911NF-20-1-0136 (simulations), the NSF QuSEC program MPS-2326748 (quantum sensing studies), and U.S. Department of Energy via the Q-NEXT Center Grant No. DOE 1F-60579 (origami studies).
We acknowledge the use of shared facilities of the UCSB Quantum Foundry through Q-AMASE-i program (NSF DMR-1906325), the UCSB MRSEC (NSF DMR 1720256), and the Quantum Structures Facility within the UCSB California NanoSystems Institute.
A.B.J. and N.Y.Y gratefully acknowledge support from the NSF QLCI program (grant no. OMA-2016245). 
A.B.J. acknowledges support from the Gordon and Betty Moore Foundation’s EPiQS Initiative via Grant GBMF10279.
N.Y.Y. acknowledges support from the Wellcome Leap Foundation under the Q for Bio programme.
L.B.H. acknowledges support from the NSF Graduate Research Fellowship Program (DGE 2139319) and the UCSB Quantum Foundry. 
R.L. acknowledges support from the NSF NRT program under award no. 2152201.

\bibliographystyle{apsrev4-2}
\bibliography{main}

\end{document}